\documentclass[aps,prb,twocolumn]{revtex4-1}
\usepackage{amsmath}
\usepackage{amssymb}
\usepackage{graphicx}
\usepackage{multirow}

\begin{document}

\title{What is the minimal model of magnetic interactions in Fe-based superconductors?}
\author{J. K. Glasbrenner}
\affiliation{National Research Council/Code 6393, Naval Research Laboratory, Washington,
DC 20375, USA}
\author{J. P. Velev}
\affiliation{Department of Physics, Institute for Functional Nanomaterials, University of
Puerto Rico, San Juan, Puerto Rico 00931, USA}
\affiliation{Code 6393, Naval Research Laboratory, Washington, DC 20375, USA}
\author{I. I. Mazin}
\affiliation{Code 6393, Naval Research Laboratory, Washington, DC 20375, USA}
\date{\today}

\begin{abstract}
Using noncollinear first-principles calculations we perform a systematic study
of the magnetic order in several families of ferropnictides. We find a fairly
universal energy dependence on the magnetization order in all cases. Our results
confirm that a simple Heisenberg model fails to account for the energy
dependence of the magnetization in couple of ways: first a biquadratic term is
present in all cases and second the magnetic moment softens depending on the
orientation. We also find that hole doping substantially reduces the biquadratic
contribution, although the antiferromagnetic stripe state remains stable within
the whole range of doping concentrations, and thus the reported lack of the
orthorhombicity in Na-doped BaFe$_2$As$_2$ is probably due to factors other than
a sign reversal of the biquadratic term. Finally, we discovered that even with
the biquadratic term, there is a limit to the accuracy of mapping the density
functional theory energetics onto Heisenberg-type models, independent of the
range of the model.
\end{abstract}

\maketitle

\section{Introduction}

\label{section-intro}

Fe-based superconductors are only the second family, after cuprates, of known
high-$T_{c}$ superconductors (HTSC). The parent compounds of these materials
exhibit magnetic ordering at low temperatures, and in their phase diagrams the
magnetic phase is proximate to the superconducting phase. The two orders are
intimately related as the superconductivity emerges when magnetism is
suppressed, for instance, by doping. Therefore, it is generally believed that
magnetic fluctuations in these systems are the likely driver of the pairing
mechanism.\cite{chubukovreview,HKM} Magnetic order is also accompanied (with a
notable exception discussed later) by a structural phase transition, and there
are compelling arguments that this is also driven by magnetism: (1) density
functional calculations quantitatively reproduce the observed orthorhombic
distortions, including the amplitude and the counterintuitive sign, and also
reproduce a qualitatively different distortion in FeTe;\cite{mazinjohannes1} (2)
the same calculations fail to produce any distortion in the absence of
magnetism. The seemingly counterintuitive fact that the structural instability
sometimes occurs at a temperature slightly above the magnetic transition is in
fact consistent with this concept, because long range magnetic order is
sufficient, but not necessary, for breaking the global C$_{4}$ symmetry: it is
enough to unequally populate magnetic fluctuations with different
\textbf{k}-vectors. These can be described as fluctuating domain walls in an
itinerant picture\cite{mazinjohannes2} or as ``order-from-disorder'' in the
local-moment picture (see Refs.~\onlinecite{chandra,Sachdev,Kivelson}; for a
review, see Ref.~\onlinecite{mazinschmalian}). This picture is also consistent
with observations of fluctuations breaking charge\cite{Japan} and
spin\cite{Matt} $C_{4}$ symmetry locally well above the N\'{e}el temperature.

\begin{figure}
\begin{center}
\includegraphics[width=0.3\textwidth]{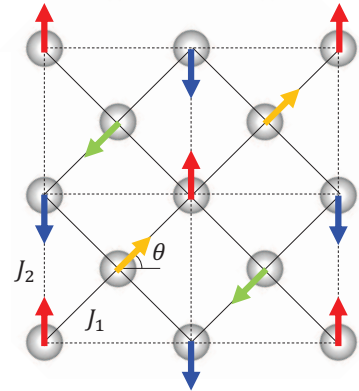}
\end{center}
\caption{(Color online) Schematic view of the two-dimensional Fe planes in the Fe-based
superconductors. The first ($J_1$) and second ($J_2$) nearest-neighbor exchange
interactions are shown in solid and dashed lines. The moments on the two
sublattices form an antiferromagnetic checkerboard pattern and in our
calculations the moments on one sublattice is rotated by an angle $\theta$
relative to the other. The antiferromagnetic stripe patterns correspond to
$\theta = 0^{\circ}$ and $\theta=180^{\circ}$.} \label{pic-schematic}
\end{figure}

The local-moment picture has the advantage of being analytically solvable and
simple; Heisenberg-like models are a popular way to approach the magnetism of
Fe-based HTSCs. This can be considered a reasonable approach since, with a
sufficient number of parameters, any sort of magnetic interaction can be mapped
onto a local moment model. The simplest possible model is a Heisenberg-type
interaction between the first and second nearest
neighbors.\cite{chandra,Sachdev,Kivelson} This model has the desired property
that symmetry breaking \textit{always} occurs above the N\'{e}el temperature,
although this splitting diminishes as the magnetic interaction becomes more
three dimensional. While the model replicates some physical properties of
Fe-based HTSCs, there is a serious problem that is often overlooked. The
essential physics of this approach can be described as
follows:\cite{mazinschmalian} an Fe plane can be viewed as a bipartite lattice
where the only interaction within each sublattice is the second nearest neighbor
exchange $J_{2}$, while the only interaction between the sublattices is $J_{1}$.
As illustrated in Fig.~\ref{pic-schematic}, exchange interaction constant $J_2$,
if $J_2 > J_1/2$, generates a checkerboard antiferromagnetic (AFM) pattern in
each sublattice, while the interaction between the sublattices cancels
completely, which holds not only for an arbitrary $J_{1}$, but for any
Heisenberg interaction of an arbitrary range. In Ref.~\onlinecite{chandra} it
was shown that, after integrating out quantum fluctuations, a nearest neighbor
biquadratic term of the form $K(\mathbf{S}_{i}\cdot \mathbf{S}_{j})$ appears in
the effective Hamiltonian, with $K>0$ and of the order 10$^{-4} J$. This lifts
the infinite degeneracy of the ground state, leaving a double degenerate state
of ferromagnetic stripes running along one of the two crystallographic
directions with AFM alternation, matching the ground state of the ferropnictides
of interest. However the small amplitude of $K$ is unphysical, making this
result purely academic and not applicable to any real material.

There is another profound problem with the Heisenberg model. Even though it
formally generates the correct ground states for ferropnictides, it fails to
explain the double stripe structure of FeTe; to do so requires introducing the
third nearest neighbor exchange $J_{3}$, which is found to be of the same order
as $J_{1}$ and $J_{2}$, a result inconsistent with the superexchange picture.
Even worse, in order to fit both the ground state and the spin wave spectra, one
needs to split the nearest neighbor exchange into two inequivalent parameters,
$J_{1a}$ and $J_{1b}$. The two parameters end up being different from each
other, sometimes even changing sign.\cite{dai} This implies that not only do the
exchange constants change qualitatively from compound to compound, but that they
have a strong, counterintuitive dependence on temperature; the inequivalent
parameters $J_{1a}$ and $J_{1b}$, above the N\'{e}el temperature, become
equivalent as required by symmetry, i.e. $J_{1a}=J_{1b}$. This bizarre and
inconsistent behavior is the death knell for using the superexchange theory to
model the Fe-based HTSCs, as there is no plausible physical mechanism that can
explain the dramatic temperature dependence of the superexchange constants.

Density functional theory (DFT) calculations, when mapped onto the same
Heisenberg model, yield similar exchange constants, including the splitting of
the nearest neighbor exchange.\cite{pickett,Antropov} This indicates that DFT is
correct in its description of the Fe-based HTSCs (it can quantitatively explain
the spin wave spectrum, for example), suggesting that this methodology can be
used to resolve the exchange constants conundrum. In fact the necessary
calculations have been reported at a very early stage, yet were largely
overlooked.\cite{yaresko} Instead of an unphysical model with only superexchange
terms, the same DFT calculations can be mapped with good accuracy onto an
isotropic ($J_{1a}=J_{1b}$) Heisenberg model which includes a biquadratic term
(formally the same as found in Ref.~\onlinecite{chandra}) with an amplitude of
$K\sim J_{1},J_{2}$. Moreover noncollinear DFT calculations with one magnetic
sublattice rotated with respect to the other, see Fig.~\ref{pic-schematic}, can
\emph{only} be mapped onto this model. It was then shown that this biquadratic,
isotropic Hamiltonian with temperature-independent parameters is an excellent
model of the magnetic properties of Fe-based HTSCs at any temperature, including
the spin wave spectra.\cite{belashchenko} The model also is consistent with an
orthorhombic transition occurring above the magnetic one.

These discoveries yielded a much more robust description of the the magnetic
behavior of the Fe-based HTSCs, which includes a consistent explanation of the
orthorhombic distortion. Instead of a minuscule ``order from disorder'' term
appearing to drive the physics of these systems, we have a sizable biquadratic
term on the mean field level. So what is missing at this point? To date, there
is no body of information about the biqudratic term. Important questions
include: how variable is it from compound to compound? How does it depend on
doping? Can it change sign, leading to a noncollinear ground state while
preserving tetragonal symmetry? At the present moment only some answers are
available. In Ref.~\onlinecite{yaresko} only two compounds were studied, and the
uncontrollable atomic spheres approximation was used. While there is no question
that the obtained results were qualitatively correct, their quantitative
accuracy remained unclear. Beyond that, in Ref.~\onlinecite{belashchenko2} it
was demonstrated that the biquadratic term depends on the details of a
material's band structure. Using accurate full potential DFT calculations with
linear muffin tin orbitals, it was shown that the biquadratic term is negative
in stoichiometric KFe$_2$Se$_2$, a hypothetical material. The term is also
dependent on the size of the local Fe moment (or, equivalently, the Fe-As/Se
bond-length distance), again indicating the necessity for accurate calculations.
Finally, the ultimate question that can be posed is whether the total energy is
mappable onto the pair interaction at all, linear or quadratic in $\mathbf{S}_i
\cdot \mathbf{S}_j$. This is always taken for granted, but there is no $a$
priori reason for that to be true in an itinerant system.

In DFT, the change in energy between different magnetic patterns accumulates via
integration over the entire occupied portion of the Fe band, which extends
several eV below the Fermi energy. Profound orbital reordering induced by
magnetism leads to the observed stripe order being lower in energy compared to
other patterns,\cite{leeku,yinleeku} and by extension affects the exchange
constants obtained by mapping. These complex changes in the electronic structure
are responsible for the anisotropy in the $J-$only model and for the large
biquadratic term in the $J-K$ model, and also for the longer range interactions
in FeTe. This contrasts with the simplistic superexchange model where both $K$
and $J_{3}$ appear only as higher order terms and must be much smaller than
$J_{1}$ and $J_{2}$.

Resolving the incomplete understanding of the biquadratic term has become even
more important after an intriguing experimental report that the orthorhombic
distortion disappears in a small part of the Ba$_{1-x}$Na$_{x}$Fe$_{2}$As$_{2}$
phase diagram, while magnetic order remains.\cite{avci} The authors favor the
plausible explanation that the biquadratic term changes sign in that region,
generating the noncollinear structure shown in Fig.~1(c) of
Ref.~\onlinecite{avci}. An alternative explanation would be that the
magnetoelastic coupling that drives the orthorhombic distortion becomes small
and the (still existing) C$_{2}$ symmetry breaking goes undetected.

The former explanation, that the biquadratic term changes sign upon doping, was
supported by the calculations of Chubukov and Eremin,\cite{ChubukovEremin} who
derived a biquadratic term in the linear response regime. The problem with this
approach is that it contradicts the DFT
finding\cite{mazinjohannes1,leeku,yinleeku} that the energy associated with
magnetic interaction accumulates over a large energy window, and that the local
moments of Fe remain large throughout the entire phase diagram. On the other
hand, knowing that magnetic ordering has a strong effect on the density of
states and the orbital composition of the Fe bands, it is plausible that
the sign of the biquadratic interaction parameter is not fixed and could change
upon doping. As previously discussed, the results of
Ref.~\onlinecite{belashchenko2} show that for the hypothetical KFe$_{2}$Se$_{2}$
compound, which can be viewed as a case of extreme hole doping, the biquadratic
term does indeed change sign.

Our goal in this paper is to make a systematic investigation of the biqudratic
interaction in representative Fe-pnictide and Fe-chalcogenide families to
address the variability of the biquadratic interaction. For this investigation,
we use the all-electron, full-potential Linear Augmented Plane Waves (FLAPW)
method. We find that the biquadratic parameter can vary within large limits, but
that it does not change sign in the accessible ranges of doping. We conclude, in
particular, that the observed lack of an orthorhombic distortion in the Na-doped
BaFe$_2$As$_2$ compound in Ref.~\onlinecite{avci} is likely due to the
inaccessibility of the tetragonal symmetry breaking by the experimental tools
used in that work.

\section{Methods}

\label{section-methodsresults}

\begin{table*}[tbp]
\caption{Summary of the crystallographic symmetry groups, lattice structure
(lattice constants $a$ and $c$, and fractional coordinates $z$ of the
non-magnetic planes), Fe-Fe/Fe-As(Se) bond lengths, the fitted parameters of
Eqs.~\ref{eq-modelfit1} \& \ref{eq-modelfit2}, and the energy change due to the
softening of the magnetic moment for all the studied compounds.}
\label{table-results}
\begin{tabular}{|c|c|c|c|c|c|c|c|c|c|c|}
\hline
Compound & Sym. & $a$ & $c$ & $z_{\text{As},\text{Se}}$ ($z_{\text{Li},\text{%
Na},\text{La}}$) & d$_{\text{Fe-Fe}}$ & d$_{\text{Fe-As(Se)}}$ & M(0) & K & J$_{\perp}$ & $E(0) \vert_{M(90)} - E(0) \vert_{M(0)}$ \\ 
& Group & (\AA ) & (\AA ) & (frac.) & (\AA ) & (\AA ) & ($\mu_B$) & (meV) & 
(meV) & (meV) \\ \hline
FeSe & P4/nmm & 3.803 & 6.084 & 0.2708 ($-$) & 2.689 & 2.516 & 2.72 & 9.67 & $-$ & -0.04 \\ \hline
LiFeAs & P4/nmm & 3.793 & 6.366 & 0.2365 (0.6541) & 2.682 & 2.421 & 1.85 & 9.75 & $-$ & 4.21 \\ \hline
NaFeAs & P4/nmm & 3.949 & 7.040 & 0.2028 (0.6460) & 2.793 & 2.437 & 2.18 & 15.43 & $-$ & 7.30 \\ \hline
LaFeAsO & P4/nmm & 4.037 & 8.742 & 0.1513 (0.6415) & 2.854 & 2.413 & 2.09 & 13.24 & $-$ & 4.97 \\ \hline
SrFe$_2$As$_2$ & I4/mmm & 3.930 & 12.324 & 0.3604 ($-$) & 2.779 & 2.390 & 1.94 & 10.57 & 0.79 & 2.78 \\ \hline
CaFe$_2$As$_2$ & I4/mmm & 3.896 & 11.683 & 0.3665 ($-$) & 2.755 & 2.376 & 1.83 & 9.39 & 1.33 & 1.58 \\ \hline
KFe$_2$As$_2$ & I4/mmm & 3.842 & 13.860 & 0.3525 ($-$) & 2.716 & 2.389 & 2.46 & 6.36 & 0.40 & 9.48 \\ \hline
KFe$_2$Se$_2$ & I4/mmm & 3.914 & 14.037 & 0.3434 ($-$) & 2.767 & 2.355 & 2.47 & -3.29 & -0.05 & 0.48 \\ \hline
BaFe$_2$As$_2$ & I4/mmm & 3.942 & 13.021 & 0.3545 ($-$) & 2.791 & 2.397 & 1.99 & 10.84 & 0.23 & 3.33 \\ 
Ba$_{0.9}$Na$_{0.1}$Fe$_2$As$_2$ & " & " & " & " & " & " & 1.96 & 9.85 & 0.21 & 4.04 \\ 
Ba$_{0.8}$Na$_{0.2}$Fe$_2$As$_2$ & " & " & " & " & " & " & 1.94 & 8.73 & 0.22 & 4.73 \\ 
Ba$_{0.7}$Na$_{0.3}$Fe$_2$As$_2$ & " & " & " & " & " & " & 1.93 & 7.80 & 0.26 & 5.30 \\ 
Ba$_{0.6}$Na$_{0.4}$Fe$_2$As$_2$ & " & " & " & " & " & " & 1.93 & 7.16 & 0.30 & 7.04 \\ 
Ba$_{0.5}$Na$_{0.5}$Fe$_2$As$_2$ & " & " & " & " & " & " & 1.94 & 5.56 & 0.24 & 5.37 \\ 
Ba$_{0.4}$Na$_{0.6}$Fe$_2$As$_2$ & " & " & " & " & " & " & 1.96 & 4.42 & 0.21 & 6.15 \\ 
BaFe$_{0.5}$Co$_{0.5}$As$_2$ & " & " & " & " & " & " & 1.27 & 0.94 & 0.039 & $-$ \\ \hline
\end{tabular}
\end{table*}

We investigate the magnetic order in representative compounds from different
families of the iron-based superconductors: FeSe for the 11 family; LiFeAs and
NaFeAs for the 111 family; BaFe$_{2}$As$_{2}$, BaFeCoAs$_{2}$, CaFe$_{2}$%
As$_{2}$, SrFe$_{2}$As$_{2}$, KFe$_{2}$As$_{2}$, and KFe$_{2}$Se$_{2}$ for the
122 family; and LaFeAsO for the 1111 family. The magnetic order is modeled using
the $J_1 - J_2 - K$ model Hamiltonian,
\begin{multline}
	\label{eq-j1j2k} H = J_1 \sum_{nn} \textbf{S}_i\cdot \textbf{S}_j +J_2 \sum_{nnn} \textbf{S}_i\cdot \textbf{S}_j \\ 
	-K \sum_{nn} \left(\textbf{S}_i\cdot \textbf{S}_j \right)^2.
\end{multline}
We use the experimentally determined crystal structures when available; for the
hypothetical material KFe$_{2}$Se$_{2}$ we used the lattice constants from
Ref.~\onlinecite{belashchenko2}. A summary of the symmetry groups, lattice
constants, and Fe-Fe and Fe-As(Se) bond lengths for all the materials is found
in Table \ref{table-results}. The FeAs(Se) planes are stacked along the c-axis
separated by a non-magnetic filler plane, except for the 11 compound FeSe which
consists only of FeSe planes. The Fe layers form a two-dimensional square
lattice and the ground state magnetic order has been confirmed both
experimentally\cite{cruz} and theoretically\cite{yaresko} to be AFM stripe
order, as schematically represented in Fig.~\ref{pic-schematic} (stripe order
corresponds to $\theta =0$ or $\theta=180^{\circ}$). In all cases in order to
accommodate the AFM stripe pattern we double the cell in the xy-plane
($\sqrt{2}\times \sqrt{2}$).

In order to study the biquadratic coupling we allow the angle $\theta$ between
the two Fe sublattices to vary. The angle $\theta$, as depicted in
Fig.~\ref{pic-schematic}, gradually interpolates between two equivalent stripes
states with $\mathbf{q}=(1,0)$ and $\mathbf{q}=(0,1)$. According to the
biquadratic model in Eq.~\ref{eq-j1j2k}, the angular energy dependence $\Delta
E(\theta) = E(\theta) - E(0)$ of the 11, 111, and 1111 families is predicted to
be,
\begin{align}
	\label{eq-modelfit1} \Delta E(\theta) = 4 K \sin^2 \theta,
\end{align}
The 122 family belongs to a centered symmetry group, and therefore rotating
$\theta $ by $180^{\circ}$ takes the system from one stripe pattern to another,
inequivalent one. The two differ by the stacking order along $c$ and the energy
difference is proportional to the interplanar exchange constant $J_{\perp}$. Taking
this into account, the angular energy dependence for the 122 family is,
\begin{align}
	\label{eq-modelfit2} \Delta E(\theta)_{122} = 4 K \sin^2 \theta	- 16 J_{\perp} \sin^2 \left( \frac{\theta}{2} \right).
\end{align}
This splits the degeneracy of the $\theta =0$ and $\theta = 180^{\circ} $ states by $16
J_{\perp}$.

In order to calculate the angular energy dependence we perform fully
noncollinear first-principles calculations using the ELK code.\cite{elk} ELK
implements density functional theory (DFT) within a FLAPW basis set with local
orbitals. In our calculations we use the Perdew-Burke-Ernzerhof (PBE)
exchange-correlation functional.\cite{pbe} ELK allows constrained magnetic
moment calculations where the moment direction and/or magnitude can be fixed. To
study the effect of hole doping on BaFe$_2$As$_2$, we use the virtual crystal
approximation (VCA) in the standard way, in which homogeneous doping is achieved
by replacing the Ba atoms with fictitious atoms of fractional nuclear charge
between those of Ba and Cs. The density of states (DOS) around the Fermi energy
of BaFe$_2$As$_2$ is dominated by Fe and As states, and so using VCA has the
primary effect of shifting the Fermi energy and removing a fractional number of
electrons from the valence band.

Convergence was checked as a function of the size of the k-point mesh. Different
size Monkhorst-Pack k-point grids were used for the different families of
compounds: $9\times 9\times 8$ for 11; $9\times 9\times 8$ for 111; $8\times
8\times 9$ for 122; and $6\times 6\times 4$ for 1111 respectively. Due to the
small energy differences the energy convergence criterion was set to 10$^{-7}$
Ha.

\section{Results and discussion}

\label{section-interpretation}

\begin{figure*}[tbp]
\begin{center}
\includegraphics[width=0.98\textwidth]{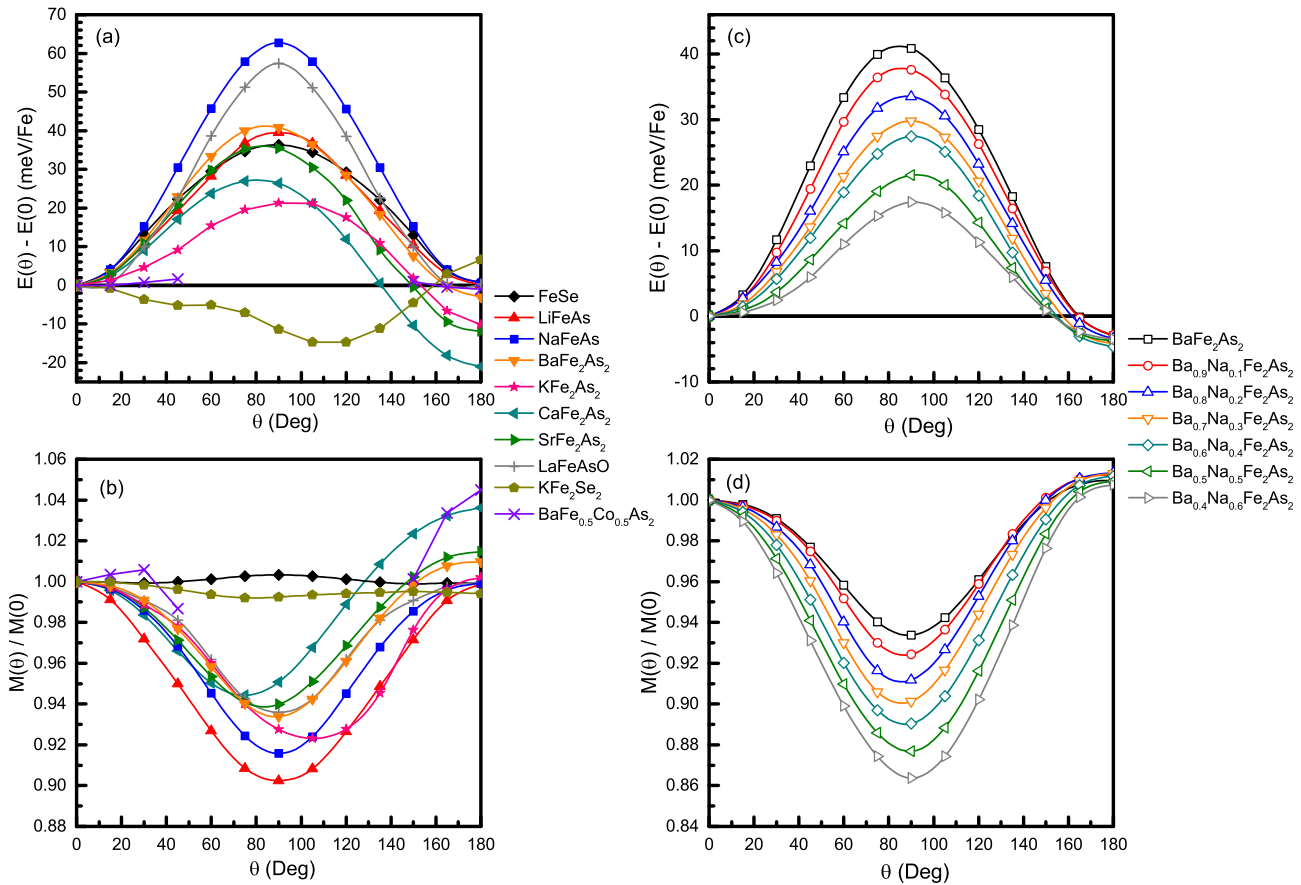}
\end{center}
\caption{(Color online) The energy $E(\protect\theta) - E(0)$ and the normalized moment $M(%
\protect\theta)/M(0)$ as a function of the relative angle $\protect\theta$
of the two magnetic sub-systems. Note that in all instances the lines are a
guide to the eye and do not represent a fit. (a) The angular dependence of
the energy for compounds belonging to the 11, 111, 122, and 1111 families of
superconductors. (b) The angular dependence of the normalized moments
belonging to the 11, 111, 122, and 1111 families of superconductors. (c) The
angular dependence of the energy for different doping levels of the 122
compound BaFe$_2$As$_2$. (d) The angular dependence of the normalized
moments for different doping levels of the 122 compound BaFe$_2$As$_2$.}
\label{pic-rotationenergy}
\end{figure*}

We checked the relative energies of the different magnetic orders. In all cases
we obtained that in the FM configuration the magnetic moment collapses and this
configuration is much higher in energy than the AFM configurations. The energy
of the checkerboard AFM configuration was found to be higher in energy than the
AFM stripe ground state; hole doping decreased the relative energy difference,
although the AFM stripe state remained as the ground state configuration.

The angular dependence of the energy difference $\Delta E(\theta )=E(\theta
)-E(0)$ for different compounds is plotted in Fig.~\ref{pic-rotationenergy}(a).
It is clear that biquadratic coupling is present in all of these compounds. We
fitted these results to Eq.~\ref{eq-modelfit1} for the 11, 111, and 1111
families and Eq.~\ref{eq-modelfit2} for the 122 family. The fitted parameters
are summarized in Table \ref{table-results}.

The angular energy dependence of the materials, with the exceptions of
KFe$_{2}$Se$_{2}$ and BaFe$_{0.5}$Co$_{0.5}$As$_{2}$, follow a similar pattern,
with the energy difference between the ground stripe state and the least
favorable $\theta = 90^{\circ}$ configuration varying between 30-60 meV/Fe. Our
results for LaFeAsO, BaFe$_{2}$As$_{2}$, and KFe$_2$Se$_2$ agree well with
previous calculations.\cite{yaresko,belashchenko2} The biquadratic interaction
constant $K$ is fairly large and positive in most materials (again excepting
KFe$_{2}$Se$_{2}$ and BaFe$_{0.5}$Co$_{0.5}$As$_{2}$) and of the same order as
$J_{1}$ and $J_{2}$.\cite{yaresko,Antropov} One of the factors that influences
$K$ is the Fe-Fe bond length, as $K$ tends to be larger in compounds which have
a greater Fe-Fe distance such as LaFeAsO, NaFeAs and BaFe$_{2}$As$_{2}$. The
interlayer coupling $J_{\perp}$ in the 122 family is about an order of magnitude
smaller than $K$, and the constant varies from material to material.

BaFe$_{0.5}$Co$_{0.5}$As$_{2}$ and KFe$_{2}$Se$_{2}$ are exceptions to the above
trends. BaFe$_{0.5}$Co$_{0.5}$As$_{2}$ was calculated using the VCA and
represents electron doping of BaFe$_{2}$As$_{2}$. This doping softens the ground
state moment by $36\%$ and destabilizes the local moment for angles $45^{\circ}
<\theta < 150^{\circ}$, in which range it collapses. This level of electron
doping also suppresses the biquadratic and interplanar interactions, reducing
both by an order of magnitude. KFe$_{2}$Se$_{2}$, on the other hand, exhibits a
negative biquadratic interaction term in agreement with the results of
Ref.~\onlinecite{belashchenko2}.\cite{footnote1} While it exhibits a negative
$K$, bulk KFe$_{2}$Se$_{2}$ is a hypothetical material that cannot be stabilized
in experiment,\cite{dingwen} so here it just serves as a proof of concept that a
negative $K$ is possible.

The softening of the moments shown in Fig.~\ref{pic-rotationenergy}(b)
contributes to $\Delta E(\theta)$ in a non-trivial way. Modeling the variation
of the magnetic moments necessitates the inclusion of $J_2$ exchange terms in
Eqs.~\ref{eq-modelfit1} \& \ref{eq-modelfit2}, as well as on-site Hund's
exchange terms such as $~IM(\theta)^2$ ($I$ is the Hund's exchange constant). A
simpler way to estimate the importance of the moment softening is to calculate
$E(0) \vert_{M(90)} - E(0) \vert_{M(0)}$ in the collinear AFM stripe state
configuration, i.e.~the energy change from reducing the ground state moment to
the self-consistent amplitude at $\theta = 90^{\circ}$, where the softening is
greatest. The final column in Table \ref{table-results} reports these
calculations. For pnictides, the difference is positive and only a few meV in
magnitude. For chalcogenides, the moment softening is slight and accordingly the
energy contribution from the softening is also small. For FeSe, the moment
amplitude grows slightly at $\theta = 90^{\circ}$, and there is a corresponding
small gain in energy. In all cases, we see that most of the energy charge in
Fig.~\ref{pic-rotationenergy}(a) is driven by the biquadratic term, with only a
modest contribution coming from the moment softening.

The angular energy dependence for various levels of hole-doping in Ba$_{1-x}$%
Na$_{x}$Fe$_{2}$As$_{2}$ via the VCA is depicted in
Fig.~\ref{pic-rotationenergy}(c). A discussion of the validity of using the VCA
to address the effect of hole-doping on the biquadratic term is included in the
Appendix. As is clear in the figure, the biquadratic interaction constant $K$
strongly depends on the degree of hole-doping in the material. Going from an
undoped system to $x=0.6$ results in a $60\%$ decrease in $K$. Extrapolating the
hole doping of BaFe$_{2}$As$_{2}$ to the extreme $x=1$ case, the biquadratic
constant $K$ however does not invert and instead nearly vanishes. Because of the
similarity of results in other materials, we expect that doping via the VCA
would yield similar results in other materials. Of course, in the case of
extreme hole doping it is necessary to allow the lattice constant and atomic
positions to relax. Here KFe$_{2}$As$_{2}$ is an example of extreme hole doping,
and if the experimental lattice constants are used, the biquadratic term is 6.3
meV, in contrast to the vanishing biquadratic term inferred from extrapolating
the VCA results discussed above. The Fe-Fe and Fe-As(Se) bond lengths,
influenced by the hole-doping level, plays a role in determining $K$. This is in
line with the results of Ref.~\onlinecite{belashchenko2}, where $K$ for
KFe$_{2}$Se$_{2}$ depended strongly on the internal coordinate $z_{\text{Se}}$.

\begin{figure}
\begin{center}
\includegraphics[width=0.48\textwidth]{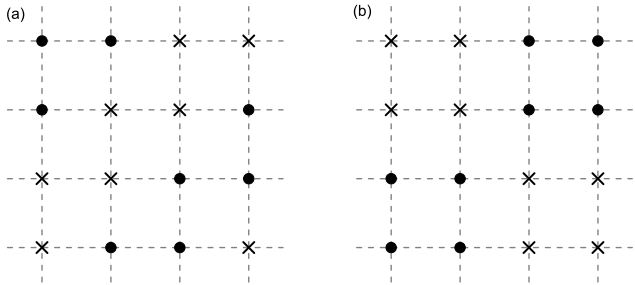}
\end{center}
\caption{The two magnetic configurations, in a $4 \times 4$ two-dimensional
cell, that are degenerate for any pairwise Hamiltonian of arbitrary range. The
closed circles correspond to spin-up moments and the crosses to spin-down
moments. (a) The double stripe configuration, which is the ground state of FeTe.
(b) The square configuration.} \label{pic-magconfigscompare}
\end{figure}

Regarding the question of whether it is valid to assume that magnetic
interactions can be accurately mapped to a pairwise Hamiltonian by just
including an arbitrary number of terms, one can address this issue by comparing
the two magnetic patterns shown in Fig.~\ref{pic-magconfigscompare}. It was
pointed out\cite{perkins} that these configurations are degenerate on the mean
field level for any Heisenberg model of arbitrary range. Being collinear, the
configurations remain degenerate after the inclusion of the biquadratic
interaction term, which can be lifted by either magnetoelastic coupling or by
integrating out fluctuations.\cite{perkins} We calculated the energy difference
between these two configurations for FeTe using the experimental high-termpature
structure (tetragonal) and found that the experimentally observed double stripe
pattern is lower in energy than the square pattern by $8$ meV/Fe, a small, but
by no means irrelevant or negligible number. The degeneracy is only lifted on
the level of the 4th order (square) ring exchange, which in the localized
Hubbard limit is of the order of $t^{4}/U^{3},$ as compared to the
nearest-neighbor superexchange terms which are of the order of $t^{2}/U.$

Another consequence of attempting to map to the classical Heisenberg
$J_1$-$J_2$-$J_3$ model is that it does not predict the double stripe or square
pattern to be the ground state of FeTe, instead predicting spiral phases for
large $J_3$. In Ref.~\onlinecite{perkins} it was pointed out that for some
parameter range collinear structures may be stabilized over the spiral ones
because of quantum fluctuations. In contrast, our calculations clearly indicate
that the sizeable biqudratic interactions, present in the parent Fe-based HTSC
compounds, completely exclude spiral phases already on the mean field level.
Moreover, DFT calculations strongly favor double stripes over squares, despite
the fact that fluctuations work in the opposite way.\cite{perkins} Our
calculations strongly suggest that the fact that the experimental structure in
FeTe appears to be double stripe is not related to magnetoelastic coupling, as
suggested in Ref.~\onlinecite{perkins}, but is instead due to itinerant effects
not captured by the Heisenberg Hamiltonian.

Returning to the experiment from Ref.~\onlinecite{avci}, we can use our results
to comment upon the experimental data. Our results do not support the proposed
noncollinear magnetic configuration in Na-doped BaFe$_2$As$_2$. Why might this
be? One possibility is that, as discussed in Sec.~\ref{section-intro}, there
might not be a reentrant C$_4$ transition in Na-doped BaFe$_2$As$_2$ and so it
would be unnecessary to argue for a change in the magnetic state. The argument
for the reentrant C$_4$ transition is based on nuclear diffractogram
measurements in which an apparent recombination of the nuclear Bragg peaks at
low temperatures is observed. However the Bragg peaks are quite broad, and so a
reduction, but not a complete removal, of the orthorhombic distortion would also
be consistent with the data. Another possibility is that the reentrant C$_4$
transition is not accompanied by a noncollinear magnetic configuration. The
authors of Ref.~\onlinecite{avci} put forth a model of the C$_4$ transition with
two magnetic configurations fitting well to measured x-ray diffraction data: the
noncollinear configuration presented in the main body of the paper and also a
collinear stripe state. The noncollinear model is preferred as it already has
C$_4$ symmetry. The authors comment that a linear combination of spin density
waves that produce stripes along the x and y directions also restores C$_4$
symmetry. Without more information, there is no reason to prefer one magnetic
configuration over the other.

Therefore, in light of our results, there remains two plausible interpretations
of Ref.~\onlinecite{avci}. The first is that the orthorhombic distortation of
Na-doped BaFe$_2$As$_2$ is reduced at lower temperatures, but ultimately retains
C$_2$ symmetry. This is the interpretation we prefer, as it is the simpler way
in which C$_2$ symmetry would be preserved. A higher resolution measurement of
the temperature dependence of the nuclear Bragg peaks is necessary to rule this
interpretation out. The second is that the C$_4$ transition does occur, but that
the magnetic order remains striped and modulates between x and y oriented stripe
patterns. Additional follow-up studies are necessary to ultimately resolve this
question.

\section{Conclusions}

\label{section-conclusion}

We confirmed that biquadratic coupling is universally present in Fe-based
superconductors. It is of the same order of magnitude as the superexchange
interactions. In the studied materials, the biquadratic term is modestly
affected by the softening of the magnetic moment, is influenced by the Fe-Fe and
Fe-As bond lengths, and depends on the doping, which underlines the biquadratic
term's itinerant origin. We find that even in the case of extreme hole doping,
no experimentally realized material exhibits a change of sign in the biquadratic
term, so the collinear AFM stripe state is energetically preferred in all
instances. Therefore, the apparent experimental observation of a reentrant C$_4$
transition in Na-doped BaFe$_2$As$_2$ is likely to be an artifact due to the
inaccessibility of measuring the C$_2$ symmetry at low temperatures with the
experimental tools.

Our results show that in the realm of Fe-based superconductors the na\"{i}ve
Heisenberg model is a rather poor approximation. The biquadratic exchange plays
an essential role and cannot be neglected in any model calculation describing
these compounds. In addition there are deviations, as observed for FeTe, from
the general pairwise interaction model for linear and biquadratic terms of an
arbitrary range. In FeTe it is these terms that stabilize the experimentally
observed double stripe in the calculations, and not the magnetoelastic coupling,
as conjectured before.\cite{perkins} It remains to be seen whether these
interesting features are specific to the parent compounds of Fe-based HTSCs or
are more common than previously expected. Further calculations and studies
should aid in answering this question.

\acknowledgments

I.I.M. acknowledges Funding from the Office of Naval Research (ONR) through the
Naval Research Laboratory's Basic Research Program. J.V. acknowledges the
support of the Office of Naval Research Summer Faculty Research Program. J.G.
acknowledges the support of the NRC program at NRL.

\appendix

\section{Validity of the virtual crystal approximation (VCA)}
\label{sect-appendix}

We confirmed the validity of the VCA by doing the following test calculations
using BaFe$_2$As$_2$. To begin, we directly substituted a Na atom for a Ba atom
(the two Ba sites are equivalent via symmetry), keeping the lattice and internal
parameters set to the values taken from experiment, and calculated $\Delta E(\pi
/2)$. For Ba$_{0.5}$Na$_{0.5}$Fe$_{2}$As$_{2}$ we obtained $\Delta E(\pi
/2)=11.4$ meV, which is about a factor of two smaller than the VCA $%
x=0.5$ result of $21.5$ meV.

To see how structural deformations affect $\Delta E(\theta)$, we relaxed the
structures in the pseudopotential-based software suite VASP.\cite{vasp1,vasp2}
In VASP we used projector augmented wave (PAW) pseudopotentials\cite{paw1,paw2}
and the Perdew-Burke-Ernzerhof generalized gradient approximation\cite{pbe} to
DFT. Our BaFe$_2$As$_2$ calculations in the main text took the internal
parameter $z_{\text{As}}$ from experiment. We want to make a proper comparison
between VCA and a relaxed structure with Na substitutions, so in order to do
this we relaxed the internal parameter $z_{\text{As}}$ for both undoped
BaFe$_2$As$_2$ and doped Ba$_{0.5}$Na$_{0.5}$Fe$_{2}$As$_{2}$ in VASP and
imported the coordinates into ELK. We then calculated $\Delta E(\pi /2)$ for
BaFe$_2$As$_2$ in the VCA with $x=0.5$ using the relaxed atomic positions found
for undoped BaFe$_2$As$_2$, and then calculated $\Delta E(\pi /2)$ for
Ba$_{0.5}$Na$_{0.5}$Fe$_{2}$As$_{2}$ using both the relaxed structure for
undoped BaFe$_2$As$_2$ and the relaxed structure for
Ba$_{0.5}$Na$_{0.5}$Fe$_{2}$As$_{2}$. The VCA result with the undoped structure
is $\Delta E(\pi /2) = 10.3 \text{ meV}$. The
Ba$_{0.5}$Na$_{0.5}$Fe$_{2}$As$_{2}$ result using the relaxed undoped structure
is $\Delta E(\pi /2) = 3.28 \text{ meV}$, and the result for using the relaxed
structure for Ba$_{0.5}$Na$_{0.5}$Fe$_{2}$As$_{2}$ is $\Delta E(\pi /2) = 7.58
\text{ meV}$. If the cell volume of Ba$_{0.5}$Na$_{0.5}$Fe$_{2}$As$_{2}$ is
fixed and both the internal parameter $z_{\text{As}}$ and $a$ and $c$ parameters
are relaxed and then imported into ELK, then $\Delta E(\pi /2) = 3.08 \text{
meV}$ for Ba$_{0.5}$Na$_{0.5}$Fe$_{2}$As$_{2}$.

Overall the VCA overestimates $\Delta E(\theta)$, but this does not affect the
qualitative behavior, i.e., the biquadratic term $K$ does not change sign.
Furthermore, using relaxed structures illustrates the sensitivity of the
biquadratic interaction to the distance between the Fe and As(Se) planes, but
these subtle changes do not materially change the overall trends. We conclude
that the VCA is an appropriate method for investigating whether doping can
affect the qualitative behavior of the biquadratic term.

\bibliography{fpnic}

\begin{thebibliography}{30}
\bibitem [{foo()}]{footnote1}%
  \BibitemOpen
  \href@noop {} {}\bibinfo {note} {In their calculations, the authors of
  Ref.~\onlinecite{belashchenko2} broke the I4/mmm symmetry and constrained the
  moments along the $c$-axis direction to be collinear in order to suppress the
  interplanar interaction term in Eq.~\ref{eq-modelfit2}. Therefore their
  $\Delta E(\theta)$ curves are symmetric about $\theta = \pi/2$ while ours are
  not.}\BibitemShut {Stop}%
\end{thebibliography}

\end{document}